\documentstyle[11pt]{article}
\begin{document}

\title{STUDY OF THE EXTINCTION LAW IN M31 AND SELECTION OF RED SUPERGIANTS}

\author{ Petko Nedialkov,\\ \\
Department of Astronomy, University of Sofia,\\
5 James Bourchier Blvd., Sofia 1164, Bulgaria\\
{\it e-mail}: japet@phys.uni-sofia.bg \\ \\
and \\ \\
Todor Veltchev,\\ \\
Institute of Astronomy, Bulgarian Academy of Sciences,\\
72 Tsarigradsko Chausse Blvd., Sofia 1784, Bulgaria\\
{\it e-mail}: eirene@phys.uni-sofia.bg }
\maketitle


\begin{abstract}
An average value of the total-to-selective-extinction ratio ${\rm R_{V}=3.8}$
${\pm 0.4}$ in M31 is obtained by means of two independent methods and by use
of the analytical formula of Cardelli, Clayton \& Mathis (1989). This result
differs from previous determinations as well from the `standard' value 3.1 for
the Milky Way. The derived individual extinctions for blue and red luminous
stars from the catalogue of Magnier {\it et al.} (1992) are in good agreement
with recent estimates for several OB associations in M31 and thus the issue
about the assumed optical opacity of the spiral disk still remains open. \\
The presented list of 113 red supergiant candidates in M31 with their
extinctions and luminosities contains 60 new objects of this type which are
not identified in other publications. It is supplemented with further 290
stars dereddened on the base of results for their closest neighbors. The
luminosity function of all red supergiant candidates and the percentage of
those with progenitors over 20 ${\rm M_{\odot}}$ suggests that the evolution
of massive stars in M31 resembles that in other Local Group galaxies.
\end{abstract}

\section{Introduction}
The interstellar dust affects considerably the statistics and the magnitudes
of the brightest stars in nearby galaxies. Its physical properties influence
the extinction law A($\lambda$)/A(V) or E($\lambda$-V)/E(B-V) which is
recently -- in most of these systems -- still poorly known due to insufficient
data. The observational derivation of this law is possible by determination of
the total-to-selective-extinction ratio ${\rm R_{V}=A(V)/E(B-V)}$ or of the
reddening line slopes in the colour-colour diagrams. Shortly before his depart
Baade said (as quoted by van den Bergh 1968) if he is allowed to choose again,
he would become an astronomer only on condition that ${\rm R_{V}}$ is
everywhere constant. There exist nowadays undoubted indications that the
average extinction law in our Galaxy depending solely on ${\rm R_{V}}$ is
applicable both to the diffuse and the dense regions of the interstellar
medium (Cardelli, Clayton \& Mathis 1989; hereafter CCM). Unfortunately the
large distances and the faint fluxes restrict the spectroscopic studies in
external galaxies -- and hence the study of ${\rm R_{V}}$ -- to observations
of high-luminosity stars.  \\
The Andromeda galaxy M31 is a favoured object for comparative investigations
and in the last years appeared new results about the extinction in it.
Combining CCD photometry and spectroscopic data for hot luminous stars, Massey
{\it et al.} (1995) (hereafter MAP95) obtained an average slope of the
reddening line ${\rm E(U-B)/E(B-V)=0.5}$ ${\pm 0.07}$ which is quite different
from the `normal' value of 0.7 in the Milky Way. Bianchi {\it et al.} (1996)
derived an ultraviolet extinction curve resembling that of the Galaxy with
possible weakness (significance of 1$\sigma$) at the 2175 $\rm\AA$ bump. In
view of these ambiguous results the knowledge of ${\rm R_{V}}$ in M31 toward
lines of sight with higher reddening is recently of great significance since: \\
1. This parameter determines the absorption corrections of the stellar
magnitude, which are crucial for clarifying the evolutionary status of the
brightest stars in M31, and it gives clue to distinguish them from the
foreground dwarfs (see the review in Massey 1998 /hereafter PM98/ and the
references therein). \\
2. It plays an important role in the transformation chain: colour excess
E(B-V) $\rightarrow$ extinction ${\rm A_{V}}$ $\rightarrow$ optical thickness
$\tau$, and thus it contributes to the study of disk's opacity.          \\
The high average metallicity Z in M31 is a well known fact (Zaritsky, Kennicutt
Jr. \& Huchra 1994). According to the recent theory of massive stars'
evolution the mass-loss rate through stellar winds increases critically with Z
(Kudritzki {\it et al.} 1989) which should lead to lowering of the upper mass
limit of the red supergiants (RSGs) (see Chiosi \& Maeder 1986). In PM98 this
prediction was supported with two facts pointed out: the lack of confirmed
RSGs of higher mass and luminosity in M31 and the metallicity-dependant red
shift of their true colours in comparison with other Local Group galaxies.
These results, however, allow an alternative explanation: a higher mean
extinction due to large amount of dust would cause poorer statistics of the
considered stars while the mentioned red shift may indicate unsufficient 
dereddening. The optical depth of the spiral disks is assumed to correlate with 
the morphological type of the galaxy and its mean surface brightness (Valentijn 
1994). Thus the disk of M31 (Sb type) would contain much more dust and should 
be less transparent in comparison with other Local Group systems studied by 
PM98. One need to mention also the limited statistics of Massey's data -- 
restricted to several OB associations -- in view of the angular size of M31 
(4x1.5 deg at the isophote ${\rm 26^{m}/arcsec^{2}}$ in B-passband). For an 
extensive survey of RSGs should be used an exclusive catalogue which covers 
the entire galaxy and contains photometry in passbands of longer effective 
wavelengths. Such requirements meets the BVRI catalogue of Magnier {\it et al.} 
(1992) (hereafter MLV) supplemented later by Haiman {\it et al.} (1994a). \\
The main goal of our work was to derive the extinction law in M31 as a first
step to obtaining a detailed absorption map of the galaxy. In the course of
this work we selected a considerable number of red stars that with high
probability do not belong to the foreground and therefore are RSG candidates.
In Section 2 we review the used observational data of MLV and discuss the
issue of the blended images. Section 3 is dedicated to the exact determination
of the parameter ${\rm R_{V}}$ as the efficiency of two alternative approaches
is critically examined. In Section 4 we present the results of dereddening for
two samples blue and red stars as well the distribution of the total
extinctions with ${\rm M_{V}}$. The list of the selected RSG candidates is
considered in Section 5 with reference to the studies of Humphreys {\it et
al.} (1988) and PM98. Its completeness and luminosity function are discussed
in Section 6. Section 7 contains a short summary of the main results.

\section{Observational Data and Their Reduction}

\subsection{The Samples and the Contamination of Foreground Stars}
Our study of the extinction in M31 is based on the large catalogue with BVRI
CCD photometry of MLV and its extension by Haiman {\it et al.} (1994a). On the
base of these data Haiman {\it et al.} (1994b) obtained extinction estimates
for 11 associations and 9 groups of associations in the eastern and western
spiral arms. The electronic version of the catalogue contains 485,388 objects
classified in each filter by DoPHOT procedure (Schechter, Mateo \& Saha 1993)
according to the characteristics of their profile. In view of our task to
perform correct dereddening in colour-colour diagrams we composed two samples
with BVI and with VRI photometry and one larger sample with VI photometry for
selecting a larger list of RSG candidates. Preliminary restriction of the
number of blended images and nonstellar objects was achieved with the
requirement each image to be classified as a star in at least one filter and
to have neither extended nor double profile in the other two. This procedure
does not solve completely the problem with blending to which we turn again in
the next subsection. \\
The contamination from Galactic dwarfs may play significant role in samples of
stars supposed to belong to large-angular-size extragalactic objects.
Predictions about the percentage of foreground stars for given bin size in
colour-colour and colour-magnitude diagrams could be made on the base of
extensive models of the Milky Way. Such a model was recently developed at the
Observatory of Besancon, France (Robin 1994). It is available in electronic
form and gives opportunity for fine tuning of magnitude, colour bin sizes and
other parameters (photometric errors, angular size of the object etc.), unlike
the traditionally used model of Ratnatunga \& Bahcall (1985). The simulations'
outcome turned out to be sensitive to the adopted fits of photometric errors
${\sigma_{B}}$ and ${\sigma_{V}}$ as functions of magnitude. Therefore we
optimized gradually our fits testing at each step the predicted numbers in
different bins of diagram V vs. (B-V) with the real counts of foreground stars
in a nearby field of M31 (Berkhuijsen {\it et al.} 1988). Reaching
satisfactory agreement, we obtained model results for diagrams (B-V) vs. (V-I)
and V vs. (V-I) where we plotted the stars from our samples. About $5 \%$ of
the subsample of 1231 blue stars selected for dereddening in colour-colour
diagram could be dwarfs from our Galaxy (see Section 4 for details). The
requirements for building a list of RSG candidates with VRI photometry and
dereddened colours practically exclude the foreground stars. The group of RSG
candidates with VI photometry suffers probably from more significant
contamination in the ranges ${\rm V<18.}$, ${\rm 2.0<(V-I)<2.2}$. \\
Identifying all of the red stars in our samples with the lists of PM98, we
found a systematic difference in V as illustrated in Figure 1a. There are
serious reasons to believe that the photometry of Massey is more accurate.
His observations were obtained with the 2.1 m Kitt Peak telescope and their
resolution is, of course, better than that of MLV. On the other side, the V
magnitudes in PM98 are in excellent agreement with the Hubble-Space-Telescope
(HST) photometry of Hunter {\it et al.} (1996) (hereafter HBOL) whereas the
values of MLV are obviously lower than the real ones (see Table 1). The latter
effect may be recognized also in diagram (V-R) vs. (V-I) where the narrow area
with highest density of red stars turned out to be parallely shifted from the
zero-absorption line toward `negative excesses'. Therefore we added to the V
magnitudes of our red stars a constant correction ${\rm \Delta_{V}=0.22}$ --
an averaged value derived from the less-squares fit for the points in Figure
1a.

\subsection{The Problem of Blended Images}
Certainly, any ground-based stellar photometry in the field of M31 obtained
with a telescope of moderate resolution could be influenced by severe blending
in the crowded regions of OB associations. HBOL warn for this effect and
illustrate it with a single example: within radius of 1 arcsec HST detected
26 stars with total magnitude close to an earlier estimate for individual
`stellar' object but the brightest component was ${\rm 2^{m}}$ fainter.
Comparing high-resolution HST data and images obtained with 1.2 and 1.3 m
telescopes, Mochejska {\it et al.} (1999) found significant contributions of
unresolved luminous companions ($19 \%$ on the average) to the measured fluxes
of the Cepheids. Since MLV performed their observations with one of those
telescopes one should expect that their catalogue is also seriously affected
by blending. Judging from the left cutoff of the FWHM's distribution (see
Figure 2 in MLV), that effect takes place for sizes well below 2.5 arcsec and
all multiple groups below 1.5 arcsec are unresolved.     
Haiman {\it et al.} (1994b) pointed out blending and differential reddening as
possible reasons for widening of the Main sequence in colour-magnitude
diagrams but, interestingly, they failed to explain with blending the
considerable spread of the stellar objects in colour-colour diagrams. Finding
the MLV data not accurate enough and perhaps assuming a lot of photometric
errors, they give up to use the Q-method and derive the extinction values
after study of the colour-magnitude diagrams for the eleven richest
associations. The noticeable number of stars with positions in colour-colour
diagram over the reddening vector for the bluest spectral type O5 was
interpreted with variety of extinction laws (different ${\rm R_{V}}$'s). That
is the case even with OB 78 -- the positions of many objects correspond to
${\rm R_{V}>6.3}$ (see Figure 8 in Haiman {\it et al.} 1994b; there
OB 013=OB 78) whereas the recent study of Veltchev, Nedialkov \& Ivanov (1999)
reveals small gas density and vanishing internal extinction probably due
to exhaustion of the natal cloud material. \\
OB 78 (NGC 206) contains numerous well crowded stars of high luminosity. Thus
the HST photometry of HBOL for this giant complex, juxtaposed with the work of
MLV, allows us to check the effect of blending in our extinction
determinations. The identification level of objects brighter than {\rm V=20.5}
turned out to be close to $100 \%$. The analysis shows that 49 from 100
identified stars over the completeness limit are substantially blended within 
radius of 1.5 arcsec. This size is slightly higher than 1/2 of the typical 
FWHM value in the data used (2.5 arcsec) and accounts also for possible errors 
in determining positions of the image centers.                            \\
In Figure 1b we compare the MLV magnitudes of blue and red blended objects
with the integral magnitudes of the components to which they are resolved in
the HST. The picture confirms the already discussed systematic difference in
V band of the MLV red stars' photometry and thus justifies the correction we
added to it. Considering the resolved groups in which the brightest component
dominates noticeably, we found that in $40 \%$ of the cases it is a single OB
star while a RSG dominates twice rarely. There are configurations of one
bright blue and one bright red component ($20 \%$ of the cases) as well more
complicated mixtures. Eventually, the effect of blending produces
predominantly objects of intermediate colour. That's why one finds in diagram
V vs. (V-I) of MLV a considerable amount of yellow stars whereas in
reality, as it is seen in Figure 4a of HBOL, in NGC 206 only a few exist. \\
The effect of blending becomes even more obvious in the colour-colour diagram
where the MLV photometry in B band has to be added. From Figure 2 it is clear
that clouds of blended images cover mainly areas over the reddening line for
spectral type O5 and with `normal' slope. Thereby larger values of ${\rm
R_V}$ are simulated. One possible way to diminish the percentage of blended
images is to use a reddening line which demarcates the large dense cloud of
blended images from area where singular images dominate. This approach was
applied as a part of our dereddening procedure for blue stars. \\
A measure for the contributions of fainter companions are the average
differences ${\rm \Delta_{V}}$, ${\rm \Delta_{I}}$ between the magnitudes of
the brightest dominant component and those of the blended image. We found out
that their typical values for the whole sample do not depend on the
photometric passband and amount 0.5 both in V and I filters. However, if one
composes two subsamples of blue and red `stars' the determined average ${\rm
\Delta_{I}}$ would not be the same -- respectively, 0.6 and 0.3 magnitudes.
Let us note also that both ${\rm \Delta_{V}}$ and ${\rm \Delta_{I}}$ depend
on magnitude -- the brighter is the dominant star the smaller they are. \\
Due to decrease both of the extinction and of the flux contribution from
fainter blue companions, the brightest RSGs (${\rm I\sim 17^{m}}$) are not
strongly affected by blending -- ${\rm \Delta_{I} \sim 0.15}$ even in the
crowded field of NGC 206. These trends are evident from Table 1 and Table 2
where we compare the VI photometries of MLV, PM98 and HBOL for 12 red images
(9 of them singular). Four of these objects should be Galactic dwarfs
according to the photometric criterion for distinguishing RSGs used by PM98
(a demarcation curve in diagram (B-V) vs. (V-R)). The Besancon model of the
Milky Way predicts, however, only 0.25 foreground stars per ${\rm arcmin^{2}}$
(${\rm 19.\le V\le 20.}$, ${\rm 2.\ge (V-I)\ge 1.8}$ which gives one star in the 
area covered by WFPC2 of HST (${\rm \approx 4.3~arcmin^{2}}$). Passing to the 
neighboring bin (${\rm 2.2\ge (V-I)\ge 2.}$) the contamination drops almost three 
times. In order to clarify the issue we combined the VI magnitudes from HBOL 
with R magnitude from PM98 and plotted the suspected Galactic dwarfs with their 
colours in diagram (V-R) vs. (V-I). Only one of them lays over the 
zero-absorption line for Main-sequence stars (cf. Figure 6) and thus certainly 
belongs to the foreground while the other three seem to be RSGs. Thereby an 
incorrect discrimination between the RSG candidates in nearby galaxies and 
Milky Way dwarfs could be explained with the effect of blending.

\section{Determination of the Parameter ${\rm R_{V}}$ }
A procedure of dereddening by means of the classical Q-method requires
exact knowledge of the total-to-selective-extinction ratio ${\rm R_{V}=}$
A(V)/E(B-V). Compiling data from various sources toward different lines of
sight, CCM derived an uniform average extinction law ${\rm <A(\lambda)/A_{V}>}$
in the infrared, optical and ultraviolet spectral ranges as function of the
wavelength and of ${\rm R_{V}}$ as a free parameter. The analytical formula is
applicable for diverse conditions in the interstellar medium and could be used
in absorption studies of diffuse as well of dense regions. It determines
unambivalent relation between the reddening line slopes in colour-colour
diagrams and ${\rm R_{V}}$ and thus allows observational estimate of this
parameter. Below we expose two approaches for working out this task.

\subsection{Test of the Reddening Line Slope on Different Colour-colour
Diagrams}                    
Let us denote the magnitudes measured in five different filters with
${\rm m_{1}}$, ${\rm m_{2}}$, ${\rm m_{3}}$, ${\rm m_{4}}$, ${\rm m_{5}}$, and
the colour excesses -- with ${\rm E(m_{1}-m_{2})}$, ${\rm E(m_{2}-m_{3})}$,
${\rm E(m_{3}-m_{4})}$, ${\rm E(m_{3}-m_{5})}$. The extinction law derived by
CCM gives the reddening line slopes
${\rm \kappa_{123}=E(m_{1}-m_{2})~/~E(m_{2}~-~m_{3})}$,
${\rm \kappa_{234}=E(m_{2}-m_{3})/}$ ${\rm E(m_{3}~-~m_{4})}$,
${\rm \kappa_{235}=E(m_{2}-m_{3})~/~E(m_{3}~-~m_{5})}$
for every value of ${\rm R_{V}}$ and thus fixes the relations between these
quantities. Naturally, two questions are rising: "Are these relations in
accord with the empirical ones which may be obtained on base of the used
stellar photometry? If yes, is this true for all possible values of ${\rm
R_V}$?" To find any answers we composed subsamples of blue stars from MLV
which belong to two distinctive associations OB48 and OB78 and are identified
with objects studied by Massey, Armandroff \& Conti (1986). The identification
provided independent UBV photometry making the obtained empirical relations
${\rm \kappa_{UBV}-\kappa_{BVR}}$ and ${\rm \kappa_{UBV}-\kappa_{BVI}}$ more
reliable. \\
Our method consists of following steps.            \\
(1) Through the zero-absorption point of the earliest type stars (O5) in
diagram (U-B) vs. (B-V) one draws a line with test value of the slope ${\rm
\kappa_{UBV}}$ corresponding to particular value of ${\rm R_V}$.    \\
(2) The identified stars within 1$\sigma$ (the individual photometric
error) from this line are selected.                    \\
(3) The selected stars are plotted with their MLV photometry in diagrams
(B-V) vs. (V-R) and (B-V) vs. (V-I) and one derives their less-squares fits
through the zero-absorption points for O5-type stars.                   \\
(4) The values of the fits' slopes are assigned respectively to ${\rm
\kappa_{BVR}}$ and ${\rm \kappa_{BVI}}$.               \\ \\
This procedure was repeated for different test slopes (different
${\rm R_{V}}$'s) and thus the desired empirical relations ${\rm \kappa_{UBV}}$
${\rm -\kappa_{BVR}}$ and ${\rm \kappa_{UBV}-\kappa_{BVI}}$ were established.
The results for OB48 and OB78 are plotted in Figure 3. One immediately sees
that the obtained ${\rm \kappa_{BVR}}$ and ${\rm \kappa_{BVR}}$ do not vary
substantially with the test slopes which perhaps reflects the lower precision
of the MLV photometry. The value of ${\rm \kappa_{UBV}}$, that corresponds to
the intersection point of the empirical curve and that derived by CCM, should
be adopted as the most probable one. It is quite the same for both
colour-colour diagrams and in both associations and gives on the average
${\rm R_{V}=3.8}$ ${\pm 0.3}$. The correctness of this estimate and its
applicability for the entire galaxy were tested by use of another approach.

\subsection{A Spectroscopy-based Approach}
Combining stellar photometry with knowledge of individual spectra, one is able
to obtain estimates of the reddening slopes ${\rm \kappa}$ in colour-colour
diagrams and hence -- following CCM -- of the individual ${\rm R_{V}}$'s. For
this purpose we identified stars from the spectroscopic studies of Humphreys,
Massey \& Freedman (1990), MAP95 and PM98, with stars from our large sample
with VI photometry. The procedure provided 18 objects of different spectral
classes and 10 of them have a measured magnitude also in R filter. We derived
their `mixed' reddening slopes ${\rm \kappa_{BVR}}$ and ${\rm \kappa_{BVI}}$
using the more accurate BV photometry of Massey, Armandroff \& Conti (1986)
and PM98, and the RI magnitudes from MLV. Of course, the errors of photometry
could yield significantly different individual slope values and therefore
${\rm \sigma_{\kappa}}$ was calculated with varying the two colours within
1$\sigma$. The zero-absorption lines for supergiants were adopted from Bessell
(1990) but their blue ends needed additional smoothing since the table points
for early spectral types refer to different luminosity subclasses. Therefore
we plotted several tracks of the Padova group (Fagotto {\it et al.} 1994) for
Z=0.5 and for more massive stars (${\rm >15~M_{\odot}}$) which were
transformed for colour-colour diagrams using the newest calibrations of
Bessell, Castelli \& Plez (1998). The blue post-Main-sequence parts of the
tracks almost coincide which gives opportunity for correct fitting of the
zero-absorption line. One should mention also their low metallicity dependence
and thus the applicability of the derived fit to the conditions in M31 (mean
Z $\sim$ 0.04). \\
We have divided the identified objects in two subsamples: early-type (A-F) and
red (K-M) stars. For particular wavelength ${\rm \lambda^{-1}}$ were retained
only those stars which give a plausible (positive) slope with not a large
error. Due to the small number of objects remaining it is appropriate to adopt
the median value ${\kappa_{med}}$ as the average one for the corresponding
group whereas the average value of ${\rm \kappa_{UBV}}$ is taken directly from
MAP95. In Figure 4 the results are compared with the curves derived by CCM for
fixed values of ${\rm R_V}$. The picture illustrates the range of possible
total-to-selective-extinction ratios which obviously embraces the estimate
from the previous subsection (solid line). In order to derive ${\rm R_V}$ more
accurately, we varied this parameter with step 0.1 and performed ${\rm
\chi^{2}}$ minimization of the individual slopes' deviations from the
extinction law of CCM. As seen from Figure 5, this method gives exactly and
unambiguously ${\rm R_{V}=3.8}$ which is also confirmed by minimization of the
deviations of ${\rm \kappa_{med}}$ for both subsamples and filter wavelengths.
One gets some idea about the uncertainties of this approach by applying the
procedure to the deviations of the {\it arithmetic averaged} values ${\rm
\kappa_{aa}}$. Finding the corresponding ${\rm R_{V}=3.4}$, we will adopt
${\rm R_{V}=3.8 \pm 0.4}$ as fixed total-to-selective-extinction ratio in all
regions of M31.

\section{Dereddening of Blue Stars and RSG Candidates}
The typical completeness limit ${\rm V_{lim}=20.7}$ of the blue stars' sample
was obtained comparing the apparent luminosity function with that of the
identified stars from Massey, Armandroff \& Conti (1986). An analogical
approach with reference to objects from PM98 leads to ${\rm V_{lim}=19.5}$
for our sample of red stars. Therefore only stars with ${\rm V\le 20.0}$ were
selected for dereddening -- the more so as the list of fainter blue and red
objects suffers both from blending and from significant foreground
contamination. This selection restricts our absorption study to the younger
massive population of M31 (top-of-the-Main-sequence stars and supergiants)
which may have some intrinsic extinction due to existing envelopes.
Preliminary differential dereddening for the Milky Way was performed for each
star of the samples with the `standard' value of ${\rm R_{V}=3.1}$. In this
procedure the excess E(B-V) was determined as function of the position in
field of M31 using the average Galactic reddening estimates of Burstein \&
Heiles (1984) for nearby areas on the sky. Its value varies from 0.06 at the
southern edge and 0.10 at the northern edge of the galaxy.

\subsection{The Standard Q-method}
Magnier {\it et al.} (1997) show that the adoption of single constant
extinction value is a too rough approximation even within M31 OB associations
where ${\rm A_V}$ varies in large ranges. Nevertheless, Haiman {\it et al.}
(1994b) do not perform differential dereddening with the MLV data
overestimating, in our view, the significance of the photometric errors. As we
have demonstrated in previous section, the proper treatment of the blending
effect still enables the classical approach of differential extinction
derivation. \\
By means of the Q-method 1231 blue stars plotted on diagram (B-V) vs. (V-I)
were dereddened individually to the zero-absorption lines for supergiants
and dwarfs of Bessell (1990). Then, comparing the assumed positions of each
star in colour-luminosity diagram in both cases, we elaborated a simple
procedure for distinguishing the Main-sequence members and for adopting of
appropriate extinction values. Stars under the reddening line for spectral
type A5 (with slope corresponding to the adopted value of ${\rm R_{V}=3.8}$)
and over the demarcation line between the clouds of singular and blended
images (see Figure 2) were preliminarily excluded from the sample. Juxtaposing
the stellar counts in diagram (B-V) vs. (V-I) with predictions of the Besancon
model of the Galaxy, one finds possible foreground contamination in several
bins which amounts ${ \sim 5 \%}$ of the whole sample. We did not take into
consideration objects with position under the `red tail' of the
zero-absorption line in order to avoid ambiguity in some dereddening
estimates. Thereby an upper limit for the calculated extinctions was set. To
diminish the number of fake blue stars, we excluded also dereddened objects
with positions in diagram ${\rm M_{V}}$ vs. ${\rm (V-I)_{0}}$ more than
1$\sigma$ over the track for 40 ${\rm M_{\odot}}$. Stars with `negative
excesses' were retained in the sample except those with unplausibly blue
colours (e.g. ${(B-V)<-0.5}$). \\
We obtained also the extinction for 113 red stars applying the Q-method in
diagram (V-R) vs. (V-I). The initial sample was composed by selection of red
objects with ${\rm V>17.}$ which lay under and in 1$\sigma$ around the
zero-absorption line for supergiants. Their number was restricted further by
the leftmost possible reddening line (see Figure 6) which determines the
spectral types to be no earlier than K6. The true colours of K6 were assigned
also to objects which lay within 1$\sigma$ to the left of that line. After
applying the Besancon model to diagram (V-R) vs. (V-I), it turned out that our
sample could be considered free from foreground contamination. The positioning
of identified stars from PM98 in this diagram (Figure 6) provides another
check of the latter suggestion. The magnitudes in V and R filters of PM98 were
supplemented with the magnitude in I band from MLV. Although in the area of
earlier spectral classes the Galactic dwarfs and RSGs of M31 (as classified by
Massey) are well mixed, one may see that is not the case with our sample.
Several foreground stars lay close and below the zero-absorption line for
supergiants but within the range of photometric errors.\\
The absolute magnitude ${\rm M_V}$ of all dereddened stars was calculated with
distance modulus 24.47 (Stanek \& Garnavich 1998). The correlation between
${\rm M_V}$ and the derived total extinction ${\rm A_V}$ is shown in Figure 7.
The well pronounced upper boundary corresponds to the limits of photometry.
The distributions for both blue and red samples are practically the same but
the former one exhibits much larger dispersion toward its more luminous tail.
That points probably to the variety of ages and respectively to different
conditions in the circumstellar environment of the massive blue stars.

\subsection{The `Closest-neighbors approach'}
In view of our goal to select more RSG candidates in M31 with known luminosity
we composed for dereddening a larger sample of 839 stars with VI photometry
only. Their magnitudes and colours were in range ${\rm 17. \le V \le 20.}$,
${\rm 2.\le (V-I)\le 3.}$ which ensures a vanishing percentage of Galactic dwarfs
(with the only exception in the bin ${\rm V>18.}$, ${\rm (V-I)<2.2}$, where it 
reaches $25 \%$). We looked up to find close neighbors among the dereddened blue 
stars for each object of this sample. Obtaining lists of 98 stars with neighbor(s)
within radius of 5 arcsec and 192 stars with neighbor(s) within radius of 10
arcsec, we assigned to them the median extinction value of their neighbors.
This estimate makes sense since such angular distances correspond to sizes
twice less than the mean size of Andromeda's OB associations (Ivanov 1996)
to which most of our blue stars belong.

\section{Selection of RSG Candidates in M31}
\subsection{Previous Photometric and Spectroscopic Studies}
The most eminent results in search for RSGs in Andromeda galaxy were achieved
in the last decade by Humphreys {\it et al.} (1988) and PM98. The former work
is based mainly on photographic photometry of Berkhuijsen {\it et al.} (1988)
and contains lists of 42 Galactic dwarfs and 23 late-type supergiants selected
through spectroscopic criteria and an analysis of diagram (J-H) vs. (H-K).
Since the completeness limits are as high as V$\approx$~18.8, R$\approx$~17.8,
that study concerns only the brighter RSG candidates. A deeper survey of red
stars in M31 presented Nedialkov, Kourtev \& Ivanov (1989) who used B and V
plates obtained with an 2 m RCC telescope and composed their sample (46 stars)
by blinking and selecting of objects with ${(B-V)>1.8}$. Their results support
the claim about lack of bright RSGs in this galaxy and especially in its
southwestern part. This statement was repeated again by PM98. Using BVR CCD
photometry on a 2.1 m telescope and plotting the objects in the colour-colour
diagram, he distinguished successfully 102 RSG candidates from 99 foreground
stars. Ten RSGs are confirmed also spectroscopically and exhibit spectral
classes from K5 to M2.5. The completeness limit of V=20.5 enables the
visibility of M supergiants with absolute magnitudes ${\rm M_{V} \approx -4.5}$
and thus makes the catalogue of PM98 an appropriate reference frame for our
selection.

\subsection{Our Lists of RSG Candidates}
In Table 3 we have put all dereddened by means of the Q-method RSG candidates
with their absolute magnitudes, true colours, extinctions and identifications,
in comparison with the works already cited. The first overview shows that we
propose 60 new candidates for RSGs which could not be found in the previously
refered works. Among the identified ones it is worth to note R-23, R-53 and
R-95 from the catalogue of Humphreys {\it et al.} (1988) whose luminosity
class is confirmed spectroscopically. The cross-identification with objects
from PM98 is very poor (only 2 RSGs) but that may well be explained with the
small total area of the fields observed by Massey: $2.3 \%$ of the region
covered by the VRI photometry of MLV. The most massive RSG candidates and
their relative number deserve special attention in view of the claim in PM98
that the most luminous of them originate as stars of $\sim$ 13-15 ${\rm
M_{\odot}}$. We composed diagram ${\rm M_V}$ vs. ${\rm (V-I)_0}$ with tracks
of the Padova group (Fagotto {\it et al.} 1994) (transformed to colours, as
already mentioned, with the calibrations of Bessell, Castelli \& Plez (1998))
and selected all dereddened stars over the track with initial mass ${\rm
20~M_{\odot}}$ as illustrated in Figure 8. Since the extinction of some of
them is derived by means of the `closest neighbors-approach' its value --
and hence the derived mass range and luminosity -- could be incorrect.
Therefore we included in this list only evolved objects with spectral class
M (${\rm (V-I)_{0}>1.8}$). The selected 68 RSG candidates are listed in
Table 4. Because of their positions in diagram V vs. (V-I) (${\rm (V-I)>2.2}$;
see Section 2.2) one can be sure none of them is a foreground dwarf.

\section{Discussion}
The full list of selected RSG candidates with derived ${\rm A_V}$ suffers
significant incompleteness even in the range of largest luminosities. The
reason lies in the problems MLV obviously had had with the photometry in R
filter. A half of the red stars identified with RSGs from PM98 and with
typical values of (V-I) have a measured magnitude in R which corresponds
to negative (V-R) colours. That restricted, of course, the number of RSG
candidates to be dereddened by means of the Q-method and diminished indirectly
the effectiveness of the `closest-neighbors approach'. A convincing
illustration of this fact provides Figure 9 where we juxtapose the luminosity
function of all dereddened red stars (solid line) with that of the 839 RSG
candidates with VI photometry (dashed line). We ascribed an average value ${\rm
A_{V}=1.0}$ to objects whose extinction we were not not able to derive
individually -- an estimate, which is close to the median values for several
OB associations studied by Magnier et al. (1997). A single look at Figure 9
reveals that in most of the bins we have `lost' about $50 \%$ of the RSG
candidates because of lack of a close neighbor. Comparing the luminosity
function of all potential RSG candidates (839 objects) with Figure 15 in PM98
and normalizing the numbers in each bin to the area of his fields, we come to
the conclusion that the completeness in our work is similar and even better in
range ${\rm M_{V}=-7.5}$ to -5.0. \\
It is worth to look intently at the shape of the obtained apparent luminosity
function of the RSGs. It differs from Massey's result for Andromeda galaxy and
resembles much more the cases in other Local Group galaxies (see Figure 15 in
PM98). The presented in PM98 distributions for M31, M33 and NGC 6822 were
interpreted there in terms of the "Conti scenario" (Conti 1976) for formation
of Wolf-Rayet stars through stellar wind -- the higher is the metallicity Z
the higher should be the mass-loss rate and the shorter time the star would
spent in the RSG stage. Hence, in galaxies with high metal abundance, one has
to expect rather sharp decrease of RSGs' number with increasing luminosity
than extended tails of the distribution. Our result for M31 casts doubt on
that assumption. We suppose that the steep luminosity function of RSGs derived
by PM98 is conditioned by applying small fixed values of colour excess E(B-V)
for each field (in the range between 0.08 and 0.25). Taking also into account
the adopted in that work lower value of ${\rm R_{V}=3.2}$, it is easy to
explain the claimed paucity of RSGs over ${\rm M_{V}=-6.0}$ and their
`concentration' in three bins down to ${\rm M_{V}=-4.0}$. We believe that the
differential dereddening we performed -- although of a limited sample -- leads
to more accurate determination of the RSGs' absolute magnitudes and reproduces
more correctly their luminosity function.              \\
Looking back at Figure 8 and Table 4, we can state with certainty that in M31
exists a significant number of RSGs whose progenitors should have masses over
${\rm 20~M_{\odot}}$. This result is in accord with the latest investigations
of Massey \& Johnson (1998) who found distinctive spatial correlation between
the luminous RSGs and Wolf-Rayet stars in this galaxy -- a fact which
testifies that stars with initial masses over ${\rm 15~M_{\odot}}$ still
undergo the phase of RSG in their evolution. The completeness limit of our
sample (cf. Figure 9), however, hinders us to estimate the relative duration
of this phase in comparison with the lifetime of RSGs of lower mass.

\section{Conclusions}
Our main result in the recent work is the derivation of a new value of the
total-to-selective-extinction ratio in M31 ${\rm R_{V}=3.8}$ ${\pm 0.4}$
obtained by means of two different methods. This estimate is considerably
higher than the previous results of Kron \& Mayall (1960) and Martin \& Shawl
(1978) but still it is within 2$\sigma$ from the `standard' value ${\rm
R_{V}=3.2}$ adopted in PM98. Nevertheless, we have reasons to infer that
the extinction law in Andromeda galaxy differs from that in the Milky Way.
Additional hints in this direction gives the work of MAP95 albeit their
calculations of the average reddening line slope lead -- applying the formula
of CCM -- to a very large and thus questionable value of ${\rm R_V}$ ($\sim$
10.). Larger values of ${\rm R_V}$ may indicate leftover remnants of molecular
clouds which surround the young stars.       \\
The obtained averaged colour excesses and total extinctions for blue and red
luminous stars in M31 are in good agreement with the newest results for
several OB associations (Magnier {\it et al.} 1997). We should point out that
most of the selected young objects belong to the spiral arms where much dust
is concentrated whereas it is hard to distinguish possible interarm stars
because of the highly inclined galactic plane. If the calculated extinctions
are corrected for the line of sight, they lead to significantly lower optical
depths than the estimates of Nedialkov (1998) for the spiral arms (${\rm
\tau_{B} \sim 4.}$). Therefore one is not allowed -- without certain
assumptions about the gas-to-star geometry -- to interpret our results as a
subsequent objection (see e.g. Disney {\it et al.} 1989, Valentijn 1990)
against the claimed optical transparency of the spirals.   \\
The selected 113 RSG candidates with accurately derived extinction and
luminosity build up the largest up-to-date list of such objects in M31. Sixty
of them are not identified in the works refered here and offer interesting
possibilities for further spectroscopic studies. We composed also two lists
of other RSG candidates with extinction calculated using the results for
their closest neighbors (within 5 and 10 arcsec). These lists contain
correspondingly 98 and 192 objects and are available in electronic form
upon request. From the three mentioned lists were extracted 68 stars with
masses presumably over 20 ${\rm M_{\odot}}$. Their relative number and the
shape of their luminosity function suggest that the massive stars' evolution
in M31 follows the same pattern as in the other Local Group galaxies without
any distinctive influence of the metallicity.

This research has made use of SIMBAD database operated at CDS, Strasbourg,
France. We are grateful to Dr. Valentin Ivanov for the helpful notes which
improved this paper. Our work is part of the project AZ. 940.4-326 of the 
Union der Deutschen Akademien der Wissenschaften and it was supported by 
grant of the Bulgarian Science Foundation.
\\ \\
{{\bf REFERENCES} \\
Berkhuijsen, E., Humphreys, R., Ghigo, F., \& Zumach, W., 1988, A\&AS, 76, 65 \\
Bessell, M., 1990, PASP, 102, 1181 \\
Bessell, M., Castelli, F., \& Plez, B., 1998, A\&A, 333, 231 \\
Bianchi, L., Clayton, G., Bohlin, R., Hutchings, J., \& Massey, P.,
     ApJ, 471, 203 \\
Burstein, D., \& Heiles, C., 1984, ApJS, 54, 33  \\
Cardelli, J., Clayton, G., \& Mathis, J., 1989, ApJ, 345, 245 (CCM) \\
Chiosi, C., \& Maeder, A., 1986, ARA\&A, 24, 329 \\
Conti, P., 1976, Mem. Soc. Roy. Sci. Liege, ${6^e}$, Ser. 9, 193 \\
Disney, M., Davies, J., \& Phillipps, S., 1989, MNRAS, 239, 939 \\
Efremov, Yu., Ivanov, G., \& Nikolov, N., 1987, Ap\&SS, 135, 119 \\
Fagotto, F., Bressan, A., Bertelli, G., \& Chiosi, C., 1994, A\&AS, 104, 365 \\
Haiman, Z., Magnier, E., Lewin, W., Lester, R., van Paradijs, J., Hasinger, G.,
     Pietsch, W., Supper, R., \& Truemper, J., 1994a, A\&A, 286, 725  \\
Haiman, Z., Magnier, E., Battinelli, P., Lewin, W., van Paradijs, J., Hasinger, G.,
     Pietsch, W., Supper, R., \& Truemper, J., 1994b, A\&A, 290, 371  \\
Humphreys, R., Massey, P., \& Freedman, W., 1990, AJ, 99, 84 \\
Humphreys, R., Pennington, R., Jones, T., \& Ghigo, F., 1988, AJ, 96, 1884 \\
Hunter, D., Baum, W., O'Neil Jr., E., \& Lynds, R., 1996, ApJ, 468, 633 (HBOL)\\
Ivanov, G., 1996, A\&A, 305, 708 \\
Kron, G., \& Mayall, N., 1960, AJ, 65, 581 \\
Kudritzki, R., Pauldrach, A., Puls, J., \& Abbot, D., 1989, A\&A, 219, 205 \\
Magnier, E., Lewin, W., van Paradijs, J., Hasinger, G., Jain, A., Pietsch, W.,
     \& Truemper, J., 1992, A\&AS, 96, 379 (MLV)  \\
Magnier, E., Hodge, P., Battinelli, P., Lewin, W., \& van Paradijs, J., 1997,
     MNRAS, 292, 490  \\
Martin, P., \& Shawl, S., 1978, ApJ, 253, 86 \\
Massey, P., 1998, ApJ, 501, 153 (PM98) \\
Massey, P., Armandroff, T., \& Conti, P., 1986, AJ, 92, 1303 \\
Massey, P., Armandroff, T., Pyke, R., Patel, K., \& Wilson, C., 1995,
     AJ, 110, 2715 (MAP95) \\
Massey, P., \& Johnson, O., 1998, ApJ, 505, 793 \\
Mochejska, B., Macri, L., Sasselov, D., Stanek, K., 1999, AJ, submitted
     (astro-ph/9908293) \\
Nedialkov, P., 1998, Ph.D. thesis, University of Sofia, Bulgaria \\
Nedialkov, P., Kourtev, R., \& Ivanov, G., 1989, Ap\&SS, 162, 1 \\
Ratnatunga, K., \& Bahcall, J., 1985, ApJS, 59, 63 \\
Robin, A., 1994, Ap\&SS, 217, 163 \\
Schechter, P., Mateo, M., \& Saha, A., 1993, PASP, 105, 1342 \\
Stanek, K., \& Garnavich, P., 1998, ApJ, 503, 131 \\
Valentijn, E., 1990, Nature, 346, 153 \\
Valentijn, E., 1994, MNRAS, 266, 614 \\
van den Bergh, S., 1968, Observatory, 88, 168 \\
Veltchev, T., Nedialkov, P., \& Ivanov, G., 1999, RevMexAA, 35, 13 \\
Zaritsky, D., Kennicutt, R., Jr., \& Huchra, J., 1994, ApJ, 420, 87 \\
}

{\bf FIGURE CAPTIONS} \\

{\bf Figure~1:} {Comparison of the MLV photometry of red stars (Y axis) with
two more recent works on M31: (a) PM98; (b) HBOL (there ${\rm V_{tot~HST}}$
denotes the integral magnitude of the components to which a `star' from the
MLV catalogue is resolved). In the latter picture the identified blue stars
are also plotted (open circles). }    

{\bf Figure~2:} {The clouds of blended images (filled circles) and single blue
objects (open squares) separated by a demarcation line (short-dashed) in
colour-colour diagram. The zero-absorption lines for supergiants (solid) and
Main-sequence stars (long-dashed) of Bessell (1990) are also drawn.}   

{\bf Figure~3:} {The empirical relations ${\rm \kappa_{UBV}-\kappa_{BVR}}$
(bottom panels) and ${\rm \kappa_{UBV}-\kappa_{BVI}}$ (upper panels) for OB48
(left panels) and OB78 (right panels) obtained on the base of photometry of
Massey, Armandroff \& Conti (1986) and MLV. The results for different test
slopes are marked with dots and connected with long-dashed line. The bars
illustrate the errors of the less-square fits. The solid curve presents the
calibrated relation of CCM while the short-dashed straight line shows the
value of ${\rm \kappa_{UBV}}$ adopted by us.}   

{\bf Figure~4:} {Values of the median slopes in colour-colour diagrams for
early-type (filled circles) and red stars (open circles) with spectroscopy
of Humphreys, Massey \& Freedman (1990), MAP95 and PM98. Three curves
calculated by use of the CCM formula are plotted for comparison -- for
${\rm R_{V}=3.0}$ (short-dashed line),for ${\rm R_{V}=6.0}$ (long-dashed line)
and for ${\rm R_{V}=3.8}$ (solid line).}   

{\bf Figure~5:} {Toward determination of ${\rm R_V}$ using ${\rm \chi^{2}}$
minimization. The solid curve represents the ${\rm \chi^{2}}$ sum of the
individual slopes' deviations from the CCM law as function of ${\rm R_V}$
(normalized arbitrarily for demonstrativeness) whereas the behaviour of this
quantity for the deviations of ${\rm \kappa_{med}}$ and ${\rm \kappa_{aa}}$
is illustrated respectively with filled and with open circles. See text.}   

{\bf Figure~6:} {Colour-colour diagram with the positions of 113 RSG
candidates (dots) which were to be dereddened by means of the Q-method. The
zero-absorption lines for supergiants (solid) and dwarfs (long-dashed) are
drawn together with the leftmost possible reddening line (short-dashed) which
corresponds to spectral class K6. The position of identified foreground stars
(open squares) and supergiants of M31 (filled squares) from PM98 are plotted
using I-band magnitudes from MLV. }

{\bf Figure~7:} {The correlation `absolute magnitude - total extinction' for
red (filled circles) and blue stars (dots) dereddened by means of the Q-method.}

{\bf Figure~8:} {The RSG candidates from our selections in diagram ${\rm M_{V}}$
vs. ${\rm (V-I)_0}$. The stellar masses in units [${\rm M_{\odot}}$] are 
indicated below the corresponding track. The RSG candidates from Table 3 are 
marked with open circles and all dereddened stars over ${\rm 20~M_{\odot}}$ 
(Table 4) -- with filled circles. The group of points with one and the same ${\rm
(V-I)_{0}=1.664}$ consists of stars to which all the true colour of spectral
type K6 was ascribed (see Section 4.1). The track of a ${\rm 30~M_{\odot}}$
star (dashed line) is calculated for lower metallicity Z=0.02 (than the typical
one for M31) and is plotted for comparison only.}

{\bf Figure~9:} {The luminosity function of all dereddened RSG candidates
(solid line) and that of the full list of 839 stars with ${\rm 17. \le V}$ ${\rm 
\le 20.}$, ${\rm 2.\le (V-I)}$ ${\le 3.}$ (dashed line). An average value 
${\rm A_{V}=1.0}$ was ascribed to objects whose extinction we were not able to 
derive individually.} 

\end{document}


\begin{deluxetable}{ccccccccccc}
\tablecaption{Singular red images from MLV and their identification in PM98 and 
HBOL. The subscripts indexes denote the work of reference. Columns 4 to 6 
contain some notes from PM98: is the star isolated (I) or it is situated in a 
crowded field (C); which is its spectral class (Sp) (if determined); is it 
classified as a RSG (rsg) or as a foreground dwarf (fgd).}
\tablewidth{0pt}
\tablehead{
${\rm \#_{MLV}}$ & 
${\rm \#_{PM98}}$ & 
${\rm \#_{HBOL}}$ & I/C & 
Sp & 
R/F & 
${\rm V_{MLV}}$ &
${\rm V_{PM98}}$ & 
${\rm V_{HBOL}}$ & 
${\rm I_{MLV}}$ & 
${\rm I_{HBOL}}$ }
\startdata
~95602&~78243 & 300048 & I &     & rsg & 19.73 & 20.05 & 20.11 & 17.78 & 17.79 \nl
~95879&~78819 & 400011 & C &K7I: & rsg & 19.42 & 19.61 & 19.66 & 17.65 & 17.71 \nl
~96624&781114 & 300043 & C &     & rsg & 19.82 & 20.23 & 20.01 & 17.41 & 17.05 \nl
~99714&~78588 & 300029 & I &     & fgd & 19.48 & 19.72 & 19.79 & 17.46 & 17.82 \nl
~98242&~78196 & 300034 & C &M1:I:& rsg & 19.70 & 19.62 & 19.85 & 17.22 & 17.47 \nl
~98447&~78728 & 300030 & C &     & rsg & 19.75 & 19.68 & 19.79 & 17.57 & 17.65 \nl
103801&~78705 & 200020 & C &     & rsg & 19.83 & 19.51 & 19.55 & 17.82 & 17.09 \nl
105354&~78191 & 200055 & I &     & rsg & 20.04 & 20.23 & 20.35 & 17.76 & 17.48 \nl
~95473&~78538 & 400016 & C & M0I & rsg & 18.72 & 19.11 & 19.92 & 16.50 & 16.63 \nl
\enddata
\end{deluxetable}

\begin{deluxetable}{ccccccccccccc}
\tablecaption{Blended red images from MLV and their indentification in PM98 and 
HBOL. The abbreviations are the same as in Table 1. The index `tot~HST'
denotes the integral magnitude of the components to which a `star' from the
MLV catalogue is resolved.}
\tablewidth{0pt}
\tablehead{
${\rm \#_{MLV}}$ & 
${\rm \#_{PM98}}$ & 
${\rm \#_{HBOL}}$ & 
I/C & 
R/F & 
${\rm V_{MLV}}$ & 
${\rm V_{PM98}}$ & 
${\rm V_{tot~HST}}$ & 
${\rm I_{MLV}}$ & 
${\rm I_{tot~HST}}$ }
\startdata
       & ~97094 &        & C &     & 19.12 &       & 19.37 & 17.33 & 17.38 \nl
       & 781111 & 300026 &   & fgd &       & 19.50 & 19.75 &       & 17.45 \nl 
       &        & 300123 &   &     &       &       & 21.03 &       & 21.18 \nl
       &        & 300566 &   &     &       &       & 22.70 &       & 22.44 \nl
       &        & 301995 &   &     &       &       & 23.80 &       & 22.44 \nl
       &        & 302603 &   &     &       &       & 24.01 &       & 22.72 \nl
       &        & 302958 &   &     &       &       & 24.12 &       & 22.69 \nl
       &        &        &   &     &       &       &       &       &       \nl
103581 &        &        & C &     & 19.23 &       & 19.71 & 16.89 & 17.26 \nl
       & ~78177 & 200032 &   & fgd &       & 20.38 & 19.87 &       & 17.32 \nl
       &        & 200735 &   &     &       &       & 22.86 &       & 22.73 \nl
       &        & 201535 &   &     &       &       & 23.50 &       & 21.87 \nl
       &        & 201951 &   &     &       &       & 23.69 &       & 23.35 \nl
       &        & 204291 &   &     &       &       & 24.34 &       & 23.01 \nl
       &        & 205765 &   &     &       &       & 24.81 &       & 21.40 \nl
       &        &        &   &     &       &       &       &       &       \nl
104157 &        &        & C &     & 19.30 &       & 19.31 & 17.00 & 17.13 \nl
       & 781125 & 200033 &   & fgd &       & 19.82 & 19.88 &       & 17.24 \nl
       &        & 200090 &   &     &       &       & 20.79 &       & 20.82 \nl
       &        & 200637 &   &     &       &       & 22.70 &       & 22.68 \nl
       &        & 200654 &   &     &       &       & 22.74 &       & 22.42 \nl
       &        & 201259 &   &     &       &       & 23.34 &       & 21.14 \nl
       &        & 201839 &   &     &       &       & 23.66 &       & 22.14 \nl
       &        & 203313 &   &     &       &       & 24.11 &       & 22.38 \nl
       &        & 204458 &   &     &       &       & 24.38 &       & 22.79 \nl
\enddata
\end{deluxetable}

\newpage
\begin{deluxetable}{cccccccccc}
\small
\tablecaption{Full list of 113 RSG candidates in M31 with extinction derived by 
means of the Q-method. In the last four columns are given the identifications 
with other works as follows: B88 -- Berkhuijsen et al. (1988), H88 -- Humphreys 
et al. (1988), Sp -- spectral classification according to H88 (if an asterisk 
follows -- according to PM98), N89 -- Nedialkov, Kourtev \& Ivanov (1989).} 
\tablewidth{0pt}
\tablehead{
\colhead{Nr. in MLV} & 
\colhead{Coordinates 2000.0} & 
\colhead{${\rm M_V}$} & 
\colhead{${\rm V}$} & 
\colhead{${\rm (V-I)_0}$} & 
\colhead{${\rm A_V}$} & 
\colhead{B88} & 
\colhead{H88} & 
\colhead{Sp} & 
\colhead{N89} }
\startdata
417577 & 11.118438~~~41.858448 & -8.437 & 18.607 & 1.972 & 2.329 &   ---   &  --- & --- &  R22  \nl
437031 & 11.638994~~~41.997456 & -8.134 & 18.068 & 1.664 & 1.482 & 41-4953 & R-23 & M0  &  R39  \nl
327891 & 11.241807~~~41.514965 & -7.631 & 19.662 & 1.784 & 2.591 &   ---   &  --- & --- &  ---  \nl
436447 & 11.634009~~~41.993141 & -7.591 & 19.063 & 2.106 & 1.934 & 41-4926 &  --- & --- &  ---  \nl
267939 & 11.132378~~~41.318951 & -7.528 & 19.682 & 1.664 & 2.515 &   ---   &  --- & --- &  ---  \nl
242943 & 10.927945~~~41.245186 & -7.476 & 19.669 & 1.664 & 2.452 &   ---   &  --- & --- &  ---  \nl
320145 & 11.160924~~~41.492786 & -7.407 & 17.396 & 2.027 & 0.102 & 41-2911 & R-95 &M1/M2&  R20  \nl
114840 & 10.644233~~~40.807350 & -7.281 & 19.080 & 1.664 & 1.682 & 40-4173 &  --- & --- &  ---  \nl
417341 & 11.512323~~~41.857262 & -7.226 & 19.191 & 1.664 & 1.702 & 41-4584 & R-53 &K5-M0&  R35  \nl
136803 & 10.543445~~~40.924122 & -7.141 & 19.452 & 1.836 & 1.910 & 40-3947 &  --- & --- &  ---  \nl
102137 & 10.196411~~~40.745743 & -7.039 & 18.970 & 2.006 & 1.331 & 40-2401 & R-30 & --- &  ---  \nl
424878 & 11.241183~~~41.901272 & -7.023 & 19.712 & 1.664 & 2.018 &   ---   &  --- & --- &  ---  \nl
229986 & 10.941835~~~41.194000 & -6.888 & 19.466 & 2.083 & 1.662 & 40-4653 &  --- & --- &  ---  \nl
417824 & 11.517426~~~41.859756 & -6.855 & 19.362 & 1.664 & 1.502 & 41-4606 &  --- & --- &  ---  \nl
~82288 & ~9.975927~~~40.667858 & -6.826 & 19.938 & 1.965 & 2.089 &   ---   &  --- & --- &  ---  \nl
226779 & 10.472175~~~41.178951 & -6.821 & 19.796 & 1.664 & 1.926 &   ---   &  --- & --- &  ---  \nl
139447 & 10.626180~~~40.940041 & -6.817 & 19.897 & 1.897 & 2.030 &   ---   &  --- & --- &  ---  \nl
285099 & 11.122199~~~41.385464 & -6.774 & 19.281 & 2.259 & 1.357 & 41-2658 & R-110& --- &  ---  \nl
380803 & 11.321597~~~41.663338 & -6.763 & 19.215 & 1.959 & 1.270 & 41-3867 &  --- & ${\rm M2.5^*}$ & --- \nl
249017 & 10.446620~~~41.260513 & -6.728 & 19.772 & 2.237 & 1.807 &   ---   &  --- & --- &  ---  \nl
223296 & 10.506104~~~41.171646 & -6.706 & 18.747 & 1.941 & 0.762 &   ---   &  --- & --- &  ---  \nl
211171 & 10.330422~~~41.143433 & -6.615 & 19.479 & 2.194 & 1.405 &   ---   &  --- & --- &  ---  \nl
179741 & 10.185312~~~41.064186 & -6.601 & 19.227 & 1.664 & 1.141 & 40-2353 &  --- & --- &  ---  \nl
206402 & 10.264723~~~41.130531 & -6.564 & 19.836 & 1.900 & 1.711 &   ---   &  --- & --- &  ---  \nl
263242 & 11.065625~~~41.297363 & -6.506 & 19.458 & 2.329 & 1.270 &   ---   &  --- & --- &  ---  \nl
132491 & 10.204449~~~40.902920 & -6.498 & 19.120 & 1.664 & 0.936 &   ---   &  --- & --- &  ---  \nl
228566 & 10.824341~~~41.187141 & -6.485 & 19.153 & 1.900 & 0.947 &   ---   &  --- & --- &  ---  \nl
232974 & 10.409659~~~41.208912 & -6.480 & 18.016 & 2.409 & 0.000 &   ---   &  --- & --- &  ---  \nl
293476 & 11.126873~~~41.409916 & -6.471 & 19.482 & 2.183 & 1.255 &   ---   & R-105& --- &  ---  \nl
432205 & 11.571004~~~41.955662 & -6.386 & 19.496 & 1.664 & 1.163 & 41-4772 &  --- & --- &  ---  \nl
286600 & 11.103818~~~41.389507 & -6.341 & 19.556 & 2.432 & 1.199 &   ---   &  --- & --- &  ---  \nl
232548 & 10.820907~~~41.206753 & -6.309 & 18.841 & 1.840 & 0.458 &   ---   &  --- & --- &  ---  \nl
158388 & 10.778870~~~41.010559 & -6.256 & 19.865 & 2.233 & 1.435 &   ---   &  --- & --- &  ---  \nl
237038 & 10.392576~~~41.226608 & -6.245 & 18.826 & 2.094 & 0.378 & 40-3411 &  --- & --- &  ---  \nl
112308 & 10.588814~~~40.792923 & -6.244 & 19.504 & 1.664 & 1.069 & 40-4062 &  --- & --- &  ---  \nl
128032 & 10.563226~~~40.876427 & -6.208 & 18.534 & 2.201 & 0.061 & 40-4007 &  --- & --- &  ---  \nl
395378 & 11.343620~~~41.753620 & -6.188 & 19.597 & 2.153 & 1.074 &   ---   &  --- & --- &  ---  \nl
103581 & 10.127163~~~40.750854 & -6.169 & 19.448 & 1.996 & 0.939 &   ---   &  --- & --- &  ---  \nl
~87586 & ~9.998288~~~40.687634 & -6.088 & 19.341 & 2.056 & 0.753 & 40-1221 &  --- & --- &  ---  \nl
140965 & 10.386338~~~40.949303 & -6.048 & 19.617 & 1.664 & 0.981 & 40-3380 &  --- & --- &  ---  \nl
222028 & 10.299899~~~41.169147 & -6.037 & 19.749 & 1.961 & 1.096 & 40-2958 &  --- & --- &  ---  \nl
325905 & 11.232757~~~41.509697 & -6.005 & 19.857 & 2.312 & 1.160 &   ---   &  --- & --- &  ---  \nl
361623 & 11.197263~~~41.607555 & -5.933 & 19.878 & 1.664 & 1.105 &   ---   &  --- & --- &  ---  \nl
~79609 & ~9.957870~~~40.653870 & -5.879 & 18.864 & 2.330 & 0.068 & 40-1036 &  --- & --- &  ---  \nl
168036 & 10.260784~~~41.035225 & -5.877 & 18.867 & 2.274 & 0.057 & 40-2744 &  --- & --- &  ---  \nl
311354 & 10.867110~~~41.469021 & -5.793 & 19.774 & 2.037 & 0.867 &   ---   &  --- & --- &  ---  \nl
289265 & 10.709546~~~41.397305 & -5.753 & 18.953 & 2.330 & 0.008 &   ---   &  --- & --- &  ---  \nl
139352 & 10.426512~~~40.939457 & -5.750 & 19.917 & 1.664 & 0.984 &   ---   &  --- & --- &  ---  \nl
~42157 & 10.158985~~~40.467342 & -5.729 & 19.551 & 1.972 & 0.610 &   ---   &  --- & --- &  ---  \nl
~83997 & 10.192136~~~40.674118 & -5.674 & 19.010 & 2.056 & 0.008 & 40-2381 &  --- & --- &  ---  \nl
407528 & 11.580252~~~41.809811 & -5.665 & 18.921 & 2.266 & 0.000 & 41-4793 &  --- & --- &  ---  \nl
~73184 & 10.237486~~~40.622486 & -5.648 & 19.313 & 2.035 & 0.287 & 40-2610 &  --- & --- &  ---  \nl
116736 & 10.082007~~~40.819996 & -5.639 & 19.654 & 2.139 & 0.613 & 40-1627 &  --- & --- &  ---  \nl
393691 & 10.935226~~~41.744274 & -5.637 & 19.133 & 2.579 & 0.059 &   ---   &  --- & --- &  ---  \nl
105384 & 10.137760~~~40.757343 & -5.617 & 19.619 & 1.813 & 0.558 &   ---   &  --- & --- &  ---  \nl
228468 & 10.940392~~~41.186668 & -5.616 & 18.942 & 2.368 & 0.000 & 40-4651 &  --- & --- &  ---  \nl
297393 & 10.881673~~~41.422993 & -5.608 & 18.862 & 2.138 & 0.000 &   ---   &  --- & --- &  ---  \nl
~66706 & ~9.936619~~~40.595116 & -5.578 & 18.892 & 2.293 & 0.000 & 40-0926 &  --- & --- &  ---  \nl
262935 & 10.561402~~~41.296246 & -5.557 & 18.913 & 2.393 & 0.000 &   ---   &  --- & --- &  ---  \nl
197557 & 10.211429~~~41.103188 & -5.548 & 19.123 & 2.163 & 0.000 &   ---   &  --- & --- &  ---  \nl
422806 & 11.153700~~~41.888088 & -5.524 & 19.614 & 1.952 & 0.422 &   ---   &  --- & --- &  ---  \nl
209021 & 10.869518~~~41.137890 & -5.522 & 19.996 & 1.664 & 0.828 &   ---   &  --- & --- &  ---  \nl
195456 & 10.403613~~~41.097816 & -5.517 & 18.932 & 2.178 & 0.240 &   ---   &  --- & --- &  ---  \nl
199087 & 10.375803~~~41.107571 & -5.511 & 19.632 & 2.224 & 0.454 &   ---   &  --- & --- &  ---  \nl
367762 & 11.166618~~~41.623421 & -5.501 & 19.556 & 1.907 & 0.350 & 41-2941 &  --- & --- &  ---  \nl
353326 & 11.382848~~~41.586491 & -5.498 & 19.184 & 2.150 & 0.000 & 41-4132 &  --- & --- &  ---  \nl
185612 & 10.709586~~~41.077473 & -5.494 & 18.976 & 2.233 & 0.000 & 40-4282 &  --- & --- &  ---  \nl
190260 & 10.606122~~~41.085457 & -5.491 & 19.746 & 2.336 & 0.549 &   ---   &  --- & --- &  ---  \nl
117794 & 10.085511~~~40.826340 & -5.488 & 19.283 & 2.363 & 0.091 &   ---   &  --- & --- &  ---  \nl
232722 & 10.406325~~~41.207687 & -5.484 & 19.276 & 2.002 & 0.068 &   ---   &  --- & --- &  ---  \nl
278708 & 11.129682~~~41.371151 & -5.481 & 19.553 & 2.106 & 0.336 &   ---   &  --- & --- &  ---  \nl
323709 & 10.772453~~~41.502758 & -5.480 & 19.885 & 1.664 & 0.663 &   ---   &  --- & --- &  ---  \nl
~97783 & 10.111373~~~40.729416 & -5.447 & 19.521 & 2.463 & 0.290 &   ---   &  --- & ${\rm M2.5^*}$& --- \nl
~85372 & 10.249909~~~40.678814 & -5.443 & 19.095 & 2.066 & 0.000 &   ---   &  --- & --- &  ---  \nl
367024 & 10.983985~~~41.621540 & -5.420 & 19.705 & 2.181 & 0.418 & 41-1966 &  --- & --- &  ---  \nl
138565 & 10.652536~~~40.934727 & -5.412 & 19.157 & 2.456 & 0.000 & 40-4189 &  --- & --- &  ---  \nl
138664 & 10.549353~~~40.935265 & -5.390 & 19.953 & 1.775 & 0.660 &   ---   &  --- & --- &  ---  \nl
~55552 & 10.285732~~~40.548210 & -5.382 & 19.342 & 2.095 & 0.051 & 40-2877 &  --- & --- &  ---  \nl
455055 & 11.441019~~~42.170864 & -5.361 & 19.109 & 2.058 & 0.000 & 41-4365 &  --- & --- &  ---  \nl
329364 & 11.282765~~~41.519047 & -5.346 & 19.833 & 2.342 & 0.476 &   ---   &  --- & --- &  ---  \nl
354873 & 11.363120~~~41.590343 & -5.342 & 19.394 & 2.176 & 0.031 & 41-4061 &  --- & --- &  ---  \nl
155336 & 10.287796~~~41.002903 & -5.331 & 19.376 & 2.243 & 0.021 & 40-2892 &  --- & --- &  ---  \nl
369008 & 11.265914~~~41.626938 & -5.315 & 19.687 & 2.679 & 0.296 & 41-3567 &  --- & --- &  ---  \nl
~99714 & 10.143638~~~40.736904 & -5.293 & 19.687 & 1.894 & 0.303 &   ---   &  --- & --- &  ---  \nl
~39305 & 10.097602~~~40.448238 & -5.287 & 19.583 & 2.470 & 0.200 & 40-1734 &  --- & --- &  ---  \nl
~90048 & 10.176888~~~40.697567 & -5.282 & 19.188 & 2.413 & 0.000 & 40-2307 &  --- & --- &  ---  \nl
207758 & 10.202614~~~41.134407 & -5.265 & 19.657 & 2.326 & 0.232 & 40-2439 &  --- & --- &  ---  \nl
375993 & 10.910624~~~41.647476 & -5.233 & 19.237 & 2.227 & 0.000 & 41-1660 &  --- & --- &  ---  \nl
~41254 & 10.032395~~~40.461670 & -5.204 & 19.340 & 2.272 & 0.000 & 40-1384 &  --- & --- &  R14  \nl
393714 & 11.218287~~~41.744453 & -5.203 & 19.593 & 2.412 & 0.085 & 41-3287 &  --- & --- &  ---  \nl
255049 & 11.021120~~~41.272854 & -5.181 & 19.786 & 1.942 & 0.273 &   ---   &  --- & --- &  ---  \nl
374507 & 11.112427~~~41.642986 & -5.165 & 19.322 & 2.486 & 0.000 & 41-2584 &  --- & --- &  ---  \nl
324841 & 11.023012~~~41.506336 & -5.160 & 19.267 & 2.306 & 0.000 &   ---   &  --- & --- &  ---  \nl
386660 & 11.047344~~~41.699619 & -5.149 & 19.504 & 2.629 & 0.000 & 41-2273 &  --- & --- &  ---  \nl
~83954 & 10.447381~~~40.673985 & -5.134 & 19.396 & 2.188 & 0.000 & 40-3643 &  --- & --- &  ---  \nl
341285 & 10.808848~~~41.553654 & -5.123 & 19.726 & 2.434 & 0.145 &   ---   &  --- & --- &  ---  \nl
315505 & 10.572876~~~41.480328 & -5.079 & 19.467 & 2.149 & 0.000 & 41-0654 &  --- & --- &  ---  \nl
185034 & 10.516809~~~41.076214 & -5.044 & 19.740 & 2.092 & 0.096 &   ---   &  --- & --- &  ---  \nl
319142 & 11.243835~~~41.490154 & -5.026 & 19.444 & 2.426 & 0.000 & 41-3440 &  --- & --- &  ---  \nl
367602 & 11.292324~~~41.622997 & -5.004 & 19.881 & 2.222 & 0.178 &   ---   &  --- & --- &  ---  \nl
131833 & 10.738815~~~40.899723 & -4.989 & 19.841 & 2.318 & 0.148 &   ---   &  --- & --- &  ---  \nl
309865 & 10.862050~~~41.465076 & -4.976 & 19.970 & 2.016 & 0.246 &   ---   &  --- & --- &  ---  \nl
282770 & 11.096257~~~41.379726 & -4.959 & 19.626 & 2.482 & 0.000 &   ---   &  --- & --- &  ---  \nl
364310 & 10.908365~~~41.614582 & -4.910 & 19.560 & 2.076 & 0.000 & 41-1646 &  --- & --- &  ---  \nl
170917 & 10.766940~~~41.043056 & -4.842 & 19.649 & 2.554 & 0.000 &   ---   &  --- & --- &  ---  \nl
268712 & 10.472554~~~41.323277 & -4.828 & 19.839 & 2.304 & 0.000 &   ---   &  --- & --- &  ---  \nl
363136 & 11.262823~~~41.611610 & -4.748 & 19.957 & 2.069 & 0.000 &   ---   &  --- & --- &  ---  \nl
402905 & 11.181502~~~41.788925 & -4.740 & 19.846 & 2.123 & 0.000 &   ---   &  --- & --- &  ---  \nl
266681 & 11.039278~~~41.311646 & -4.740 & 19.771 & 2.359 & 0.000 &   ---   &  --- & --- &  ---  \nl
234160 & 10.834925~~~41.214603 & -4.734 & 19.688 & 2.724 & 0.000 &   ---   &  --- & --- &  ---  \nl
139081 & 10.359098~~~40.937881 & -4.706 & 19.779 & 2.181 & 0.000 & 40-3255 &  --- & --- &  ---  \nl
196944 & 10.448886~~~41.101601 & -4.649 & 19.821 & 2.140 & 0.000 &   ---   &  --- & --- &  ---  \nl
102808 & 10.292471~~~40.748173 & -4.506 & 19.964 & 2.182 & 0.000 &   ---   &  --- & --- &  ---  \nl
\enddata
\end{deluxetable}

\newpage
\begin{deluxetable}{ccccccccccc}
\small
\tablecaption{RSG candidates in M31 whose progenitors should have masses over ${\rm
20 M_{\odot}}$). The abbreviations are the same as in Table 3. The 8th column
with designation `E' denotes the method of extinction derivation as follows: Q
-- Q-method, A -- `closest-neighbors-approach' within 5 arcsec, B -- `closest-
neighbors-approach' within 10 arcsec.} 
\tablewidth{0pt}
\tablehead{
Nr. in MLV & 
Coordinates 2000.0 &
${\rm M_V}$ & 
${\rm V}$ & 
${\rm (V-I)_0}$ & 
${\rm A_V}$ & 
E & 
B88 & 
H88 & 
Sp & 
N89}
\startdata
417577 & 11.118438~~~41.858448 & -8.437 & 18.607 & 1.972 & 2.329 & Q &  ---    &  ---  &  ---  &  R22 \nl
~~5370 & ~9.887635~~~40.151890 & -8.024 & 18.093 & 1.860 & 1.647 & B & 39-0998 &  ---  &  ---  &  --- \nl
~30186 & ~9.965296~~~40.380390 & -7.932 & 18.962 & 1.844 & 2.424 & A &  ---    &  ---  &  ---  &  --- \nl
130350 & 10.390935~~~40.892464 & -7.690 & 18.130 & 2.067 & 1.350 & B & 40-3404 &  ---  &  ---  &  --- \nl
436447 & 11.634009~~~41.993141 & -7.591 & 19.063 & 2.106 & 1.934 & Q & 41-4926 &  ---  &  ---  &  --- \nl
351497 & 11.399905~~~41.581776 & -7.567 & 17.464 & 1.820 & 0.561 & B & 41-4215 & R-88  &  ---  &  --- \nl
256709 & 10.394132~~~41.277287 & -7.497 & 17.682 & 2.175 & 0.709 & B & 41-0214 & R-175 &   M3  &  --- \nl
129360 & ~9.963593~~~40.884216 & -7.486 & 18.407 & 2.343 & 1.423 & B & 40-1063 & IIIR-2&  ---  &  R11 \nl
320145 & 11.160924~~~41.492786 & -7.407 & 17.396 & 2.027 & 0.102 & Q & 41-2911 & R-95  &  M1/M2&  R20 \nl
304588 & 11.113409~~~41.448463 & -7.312 & 18.241 & 2.558 & 1.082 & B &  ---    & ---   &  ---  &  --- \nl
~11684 & ~9.857494~~~40.236900 & -7.308 & 18.833 & 2.383 & 1.671 & B &  ---    & ---   &  ---  &  --- \nl
106445 & 10.247663~~~40.761742 & -7.295 & 18.472 & 2.057 & 1.297 & B & 40-2678 & IIIR-31& ---  &  --- \nl
427269 & 11.116948~~~41.917397 & -7.269 & 17.201 & 2.043 & 0.000 & B &  ---    & ---   &  ---  &  --- \nl
~33436 & ~9.391909~~~40.403561 & -7.248 & 18.012 & 2.273 & 0.790 & A &  ---    & ---   &  ---  &   R1 \nl
216250 & 10.252391~~~41.155861 & -7.247 & 18.223 & 1.806 & 1.000 & A &  ---    & ---   &  ---  &  --- \nl
439882 & 12.023583~~~42.026192 & -7.213 & 18.017 & 2.144 & 0.760 & B &  ---    & ---   &  ---  &  --- \nl
346421 & 11.278433~~~41.568855 & -7.199 & 17.720 & 1.840 & 0.449 & B & 41-3637 &R-81b  &  ---  &  --- \nl
338914 & 10.559035~~~41.546288 & -7.184 & 19.260 & 1.979 & 1.974 & A &  ---    & ---   &  ---  &  --- \nl
463349 & 11.718462~~~42.250755 & -7.144 & 18.490 & 2.415 & 1.164 & B &  ---    & ---   &  ---  &  --- \nl
136803 & 10.543445~~~40.924122 & -7.141 & 19.452 & 1.836 & 1.910 & Q & 40-3947 & ---   &  ---  &  --- \nl
102137 & 10.196411~~~40.745743 & -7.039 & 18.970 & 2.006 & 1.331 & Q & 40-2401 &IIIR-30&  ---  &  --- \nl
~~8380 & ~9.705883~~~40.199226 & -6.986 & 18.865 & 2.349 & 1.381 & A &  ---    & ---   &  ---  &  --- \nl
~95473 & 10.128913~~~40.719624 & -6.973 & 18.943 & 1.912 & 1.446 & A & 40-1980 & ---   &${\rm M0^*}$& --- \nl
229986 & 10.941835~~~41.194000 & -6.888 & 19.466 & 2.083 & 1.662 & Q & 40-4653 & ---   &  ---  &  --- \nl
358614 & 11.341266~~~41.599960 & -6.875 & 18.422 & 2.255 & 0.827 & B &  ---    & ---   &  ---  &  --- \nl
204943 & 10.787182~~~41.125546 & -6.848 & 18.061 & 2.608 & 0.439 & B & 40-4392 &R-131  &  ---  &  --- \nl
~82288 & ~9.975927~~~40.667858 & -6.826 & 19.938 & 1.965 & 2.089 & Q &  ---    & ---   &  ---  &  --- \nl
385173 & 11.519415~~~41.689789 & -6.825 & 17.761 & 2.013 & 0.116 & B & 41-4616 & ---   &  ---  &  --- \nl
286440 & 10.409627~~~41.388996 & -6.825 & 18.002 & 2.017 & 0.357 & B & 41-0246 & R-171 &  M0   &  --- \nl
139447 & 10.626180~~~40.940041 & -6.817 & 19.897 & 1.897 & 2.030 & Q &  ---    & ---   &  ---  &  --- \nl
~27200 & ~9.838799~~~40.358631 & -6.811 & 18.986 & 2.018 & 1.327 & B & 40-0528 & IIIR-99& ---  &   R8 \nl
411624 & 11.002398~~~41.827980 & -6.811 & 18.357 & 1.878 & 0.698 & A & 41-2045 & R-40  &  ---  &  --- \nl
463820 & 11.352789~~~42.255173 & -6.790 & 18.579 & 1.980 & 0.899 & B &  ---    & ---   &  ---  &  --- \nl
458999 & 11.381101~~~42.214844 & -6.784 & 18.999 & 1.811 & 1.313 & B &  ---    & ---   &  ---  &  --- \nl
285099 & 11.122199~~~41.385464 & -6.774 & 19.281 & 2.259 & 1.357 & Q & 41-2658 & R-110 &  ---  &  --- \nl
~13787 & 10.081344~~~40.256355 & -6.765 & 19.296 & 1.980 & 1.591 & B & 39-1144 & ---   &  ---  &  --- \nl
380803 & 11.321597~~~41.663338 & -6.763 & 19.215 & 1.959 & 1.270 & Q & 41-3867 & ---   &${\rm M2.5^*}$& --- \nl
153888 & 10.411914~~~40.999454 & -6.749 & 18.345 & 2.182 & 0.624 & B & 40-3506 & ---   &  ---  &  --- \nl
459075 & 11.381123~~~42.215511 & -6.733 & 19.050 & 1.821 & 1.313 & B & 41-4128 & ---   &  ---  &  --- \nl
249017 & 10.446620~~~41.260513 & -6.728 & 19.772 & 2.237 & 1.807 & Q &  ---    & ---   &  ---  &  --- \nl
223296 & 10.506104~~~41.171646 & -6.706 & 18.747 & 1.941 & 0.762 & Q &  ---    & ---   &  ---  &  --- \nl
~77174 & 10.644314~~~40.642048 & -6.689 & 19.392 & 1.953 & 1.611 & B &  ---    & ---   &  ---  &  --- \nl
306484 & 10.724109~~~41.455193 & -6.620 & 18.509 & 1.961 & 0.659 & A &  ---    & ---   &  ---  &  --- \nl
211171 & 10.330422~~~41.143433 & -6.615 & 19.479 & 2.194 & 1.405 & Q &  ---    & ---   &  ---  &  --- \nl
~99820 & 10.678908~~~40.737354 & -6.575 & 18.830 & 2.343 & 0.935 & B & 40-4234 & ---   &  ---  &  --- \nl
~~~916 & ~9.840187~~~40.074886 & -6.574 & 18.965 & 2.519 & 1.069 & B & 39-0926 & ---   &  ---  &  --- \nl
206402 & 10.264723~~~41.130531 & -6.564 & 19.836 & 1.900 & 1.711 & Q &  ---    & ---   &  ---  &  --- \nl
457329 & 11.344069~~~42.196171 & -6.536 & 19.203 & 1.914 & 1.269 & B & 41-3973 & ---   &  ---  &  --- \nl
154736 & 10.651431~~~41.001438 & -6.531 & 18.642 & 1.883 & 0.703 & A & 40-4186 & ---   &  ---  &  --- \nl
334859 & 10.554651~~~41.534081 & -6.516 & 19.928 & 1.821 & 1.974 & B &  ---    & ---   &  ---  &  --- \nl
332378 & 11.152265~~~41.527172 & -6.513 & 18.496 & 2.510 & 0.539 & B &  ---    & ---   &  ---  &  --- \nl
263242 & 11.065625~~~41.297363 & -6.506 & 19.458 & 2.329 & 1.270 & Q &  ---    & ---   &  ---  &  --- \nl
234927 & 10.979662~~~41.218048 & -6.491 & 18.067 & 2.204 & 0.088 & A & 40-4707 & ---   &  ---  &  --- \nl
228566 & 10.824341~~~41.187141 & -6.485 & 19.153 & 1.900 & 0.947 & Q &  ---    & ---   &  ---  &  --- \nl
232974 & 10.409659~~~41.208912 & -6.480 & 18.016 & 2.409 & 0.000 & Q &  ---    & ---   &  ---  &  --- \nl
293476 & 11.126873~~~41.409916 & -6.471 & 19.482 & 2.183 & 1.255 & Q &  ---    & R-105 &  ---  &  --- \nl
105041 & 10.145855~~~40.756027 & -6.461 & 19.260 & 2.251 & 1.251 & A &  ---    & ---   &  ---  &  --- \nl
~26754 & 10.227870~~~40.355621 & -6.450 & 19.694 & 2.160 & 1.674 & B & 40-2564 & ---   &  ---  &  --- \nl
209898 & 10.286233~~~41.140270 & -6.448 & 18.022 & 2.434 & 0.000 & A &  ---    & ---   &  ---  &  --- \nl
112591 & ~9.934171~~~40.794487 & -6.428 & 19.733 & 2.015 & 1.691 & B & 40-0914 & ---   &  ---  &  --- \nl
475471 & 11.534462~~~42.385521 & -6.424 & 19.905 & 1.974 & 1.859 & B &  ---    & ---   &  ---  &  --- \nl
150939 & 10.640384~~~40.992729 & -6.416 & 18.302 & 2.074 & 0.248 & A & 40-4159 & ---   &  ---  &  --- \nl
384965 & 11.306774~~~41.688457 & -6.392 & 18.579 & 2.223 & 0.501 & B & 41-3786 & ---   &  ---  &  --- \nl
~17645 & ~9.680671~~~40.292290 & -6.390 & 19.091 & 2.133 & 1.011 & B & 40-0114 & ---   &  ---  &  --- \nl
405303 & 10.993254~~~41.800545 & -6.378 & 18.780 & 1.829 & 0.688 & A & 41-2011 & R-42  &  ---  &  --- \nl
285052 & 11.028757~~~41.385342 & -6.347 & 18.859 & 1.811 & 0.736 & B & 41-2160 & ---   &  ---  &  --- \nl
232548 & 10.820907~~~41.206753 & -6.309 & 18.841 & 1.840 & 0.458 & Q &  ---    & ---   &  ---  &  --- \nl
299880 & 10.931031~~~41.431850 & -6.290 & 18.758 & 1.882 & 0.577 & B &  ---    & ---   &  ---  &  --- \nl
\enddata
\end{deluxetable}